\documentclass[fleqn,10pt]{wlscirep}
\usepackage{amssymb}
\usepackage{amsmath}
\usepackage{graphicx}
\usepackage{epsfig}
\usepackage{float}

\begin{document}

\title{Majorana flat band edge modes of topological gapless phase in 2D
	Kitaev square lattice }
\author[1]{K. L. Zhang}
\author[1]{P. Wang}
\author[1,*]{Z. Song}
\affil[1]{School of Physics, Nankai University, Tianjin 300071, China}
\affil[*]{songtc@nankai.edu.cn}

\begin{abstract}
We study a Kitaev model on a square lattice, which describes topologically
trivial superconductor when gap opens, while supports topological gapless
phase when gap closes. The degeneracy points are characterized by two
vortices in momentum space, with opposite winding numbers, which are not
removable unless meet together. We show rigorously that the topological
gapless phase always hosts a partial Majorana flat band edge modes in a
ribbon geometry, although such a single band model has zero Chern number as
a topologically trivial superconductor. The flat band disappears when the
gapless phase becomes topologically trivial, associating with the mergence
of two vortices. Numerical simulation indicates that the flat band is robust
against the disorder. This finding indicates that the bulk-edge
correspondence can be extended to superconductors in the topologically
trivial regime as recently proposed in Ref. [PRL 118, 147003 (2017)].
\end{abstract}

\maketitle

\newpage

\section*{Introduction}

Topological materials have become the focus of intense research in the last
years \cite{Hasan,XLQ,CKC,HMW}, since they not only exhibit new physical
phenomena with potential technological applications, but also provide a
fertile ground for the discovery of fermionic particles and phenomena
predicted in high-energy physics, including Majorana \cite%
{LF,RML,VM,SNP,YO,NR}, Dirac \cite{AHCN,ZKL,ZKL2,JAS,ZW,JX,SMY} and Weyl
fermions \cite{MH,SMH,BQL,BQL2,CS,XW,HW,SYX,SYX2}.\ These concepts relate to
Majorana edge modes and topological gapless phases. System in the
topological gapless phase exhibits band structures with band-touching points
in the momentum space, where these kinds of nodal points appear as
topological defects of an auxiliary vector field. Then these points are
unremovable due to the symmetry protection, until a pair of them meets and
annihilates together. On the other hand, a gapful phase can be topologically
non-trivial, commonly referred to as topological insulators and
superconductors, the band structure of which is characterized by nontrivial
topology. A particularly important concept is the bulk-edge correspondence,
which links the nontrivial topological invariant in the bulk to the
localized edge modes. The number of Majorana edge modes is determined by
bulk topological invariant. In general, edge states are the eigenstates of
Hamiltonian that are exponentially localized at the boundary of the system.
The Majorana edge modes have been actively pursued in condensed matter
physics \cite{AJ,BCWJ,STD,LM,ESR,DSS,SM} since spatially separated Majorana
fermions lead to degenerate ground states, which encode qubits immune to
local decoherence \cite{NC}. \ This bulk-edge correspondence indicates that
a single-band model must have vanishing Chern number and there should be no
edge modes when open boundary conditions are applied. However, the existence
of topological gapless indicates that there is hidden topological feature in
some single band system. A typical system is a 2D honeycomb lattice of been
graphene, which is a zero-band-gap semiconductor with a linear dispersion
near the Dirac point. Meanwhile, there is another interesting feature lies
in the appearance of partial flat band edge modes in a ribbon geometry \cite%
{FM,RS,YW}, which exhibit robustness against disorder \cite{WK}. Recently,
it has been pointed that Majorana zero modes are not only attributed to
topological superconductors. A 2D topologically trivial superconductors
without chiral edge modes can host robust Majorana zero modes in topological
defects \cite{ZBY}. All these facts stimulate a question that the bulk-edge
correspondence may be extended to a topological gapless phase.

In this paper, we investigate this issue through an exact solution of a
concrete system. We study a Kitaev model on a square lattice, which
describes topologically trivial superconductor when gap opens, while
supports topological gapless phase when gap closes \cite{WP1,WP2}. The
degeneracy points are characterized by two vortices, or Weyl nodal points in
momentum space, with opposite winding numbers, which are not removable
unless meet together. This work aims to shed light on the nature of
topological edge modes associated with topological gapless phase, rather
than gapful topological superconductor. We show rigorously that the
topological gapless phase always hosts a partial Majorana flat band edge
modes in a ribbon geometry. The flat band disappears when the gapless phase
becomes topologically trivial, associating with the mergence of two
vortices. Numerical simulation indicates that the flat band is robust
against the disorder.

\section{Results}

We have demonstrated that a topologically trivial superconductor emerges as
a topological gapless state, which support Majorana flat band edge modes.
The new quantum state is characterized by two linear band-degeneracy points
with opposite topological invariant. In sharp contrast to the conventional
topological superconductor, such a system has single band, thus has zero
Chern number. We prove that the appearance of this topological feature
attributes to the corresponding Majorana lattice structure, which is a
modified honeycomb lattice. It is natural to acquire a set of zero modes,
which is robust against disorder. The results of our model indicate that the
bulk-edge correspondence can be extended to a single-band system with hidden
topological feature. In the following, there are three parts: (i) We present
the Kitaev Hamiltonian on a square lattice and the phase diagram for the
topological gapless phase. (ii) We investigate the Majorana bound states.
(iii) We perform numerical simulation to investigate the robust of the edge
modes against the disorder perturbations.

\subsection{Model and topological gapless phase}

We consider the Kitaev model on a square lattice which is employed to depict 
$2$D $p$-wave superconductors. The Hamiltonian of the tight-binding model
reads%
\begin{eqnarray}
H &=&-t\sum_{\mathbf{r,a}}c_{\mathbf{r}}^{\dagger }c_{\mathbf{r}+\mathbf{a}%
}-\Delta \sum_{\mathbf{r},\mathbf{a}}c_{\mathbf{r}}c_{\mathbf{r}+\mathbf{a}}+%
\text{h.c.}  \notag \\
&&+\mu \sum_{\mathbf{r}}\left( 2c_{\mathbf{r}}^{\dagger }c_{\mathbf{r}%
}-1\right) ,
\end{eqnarray}%
where $\mathbf{r}$ is the coordinates of lattice sites and $c_{\mathbf{r}}$
is the fermion annihilation operators at site $\mathbf{r}$. Vectors $\mathbf{%
a}=a\mathbf{i},$ $a\mathbf{j},$ are the lattice vectors in the $x$ and $y$
directions with unitary vectors $\mathbf{i}$\ and $\mathbf{j}$. The hopping
between neighboring sites is described by the hopping amplitude $t$. The
isotropic order parameter $\Delta $ is real, which result in topologically
trivial superconductor. The last term gives the chemical potential.

\begin{figure}[h!]
\centering
\includegraphics[bb=0 0 790 410, width=1\textwidth, clip]{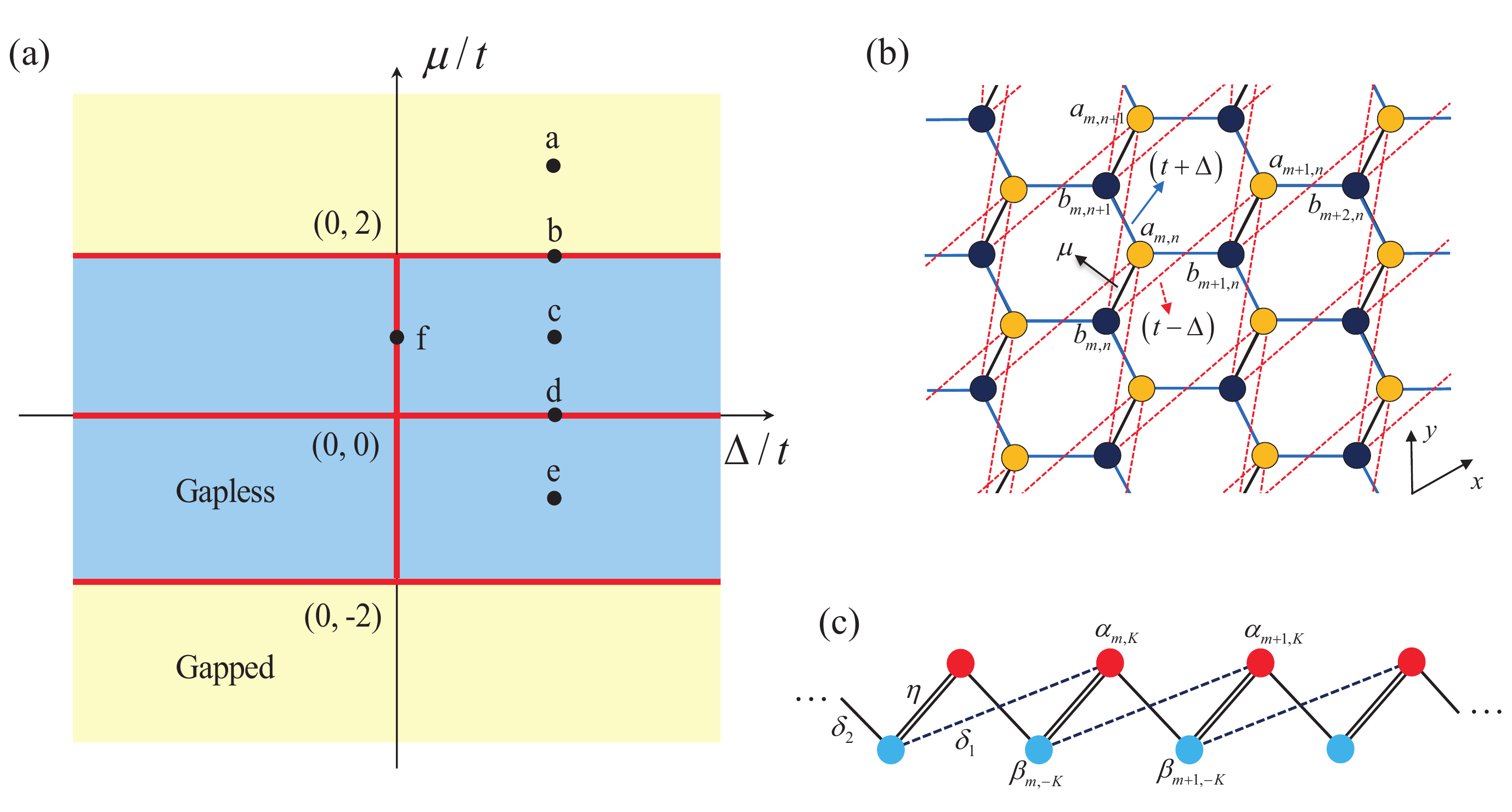} 
\caption{(Color online) (a) Phase diagram of the Kitaev model on a square
lattice system on the parameter $\protect\mu -\Delta $ plane (in units of $t$%
). The red lines indicate the boundary, which separate the topologically
trivial gapped phases (yellow) and topological gapless phases (blue). The
system at the boundary (red lines) is topologically trivial gapless phase.
(b) Schematically illustration of the Majorana lattice, which is honeycomb
geometry with long-range hopping term. (c) The geometry of the auxiliary
operator lattice represented in Eq. (\protect\ref{SSH}), which is an SSH\
chain with long-range hopping term. The edge modes of a set of modified SSH
chains form Majorana fat band edge modes in (b) when the system is in the
blue region of the phase diagram (a).}
\label{fig1}
\end{figure}

\begin{figure}[h!]
\centering
\includegraphics[bb=75 150 530 635, width=1\textwidth, clip]{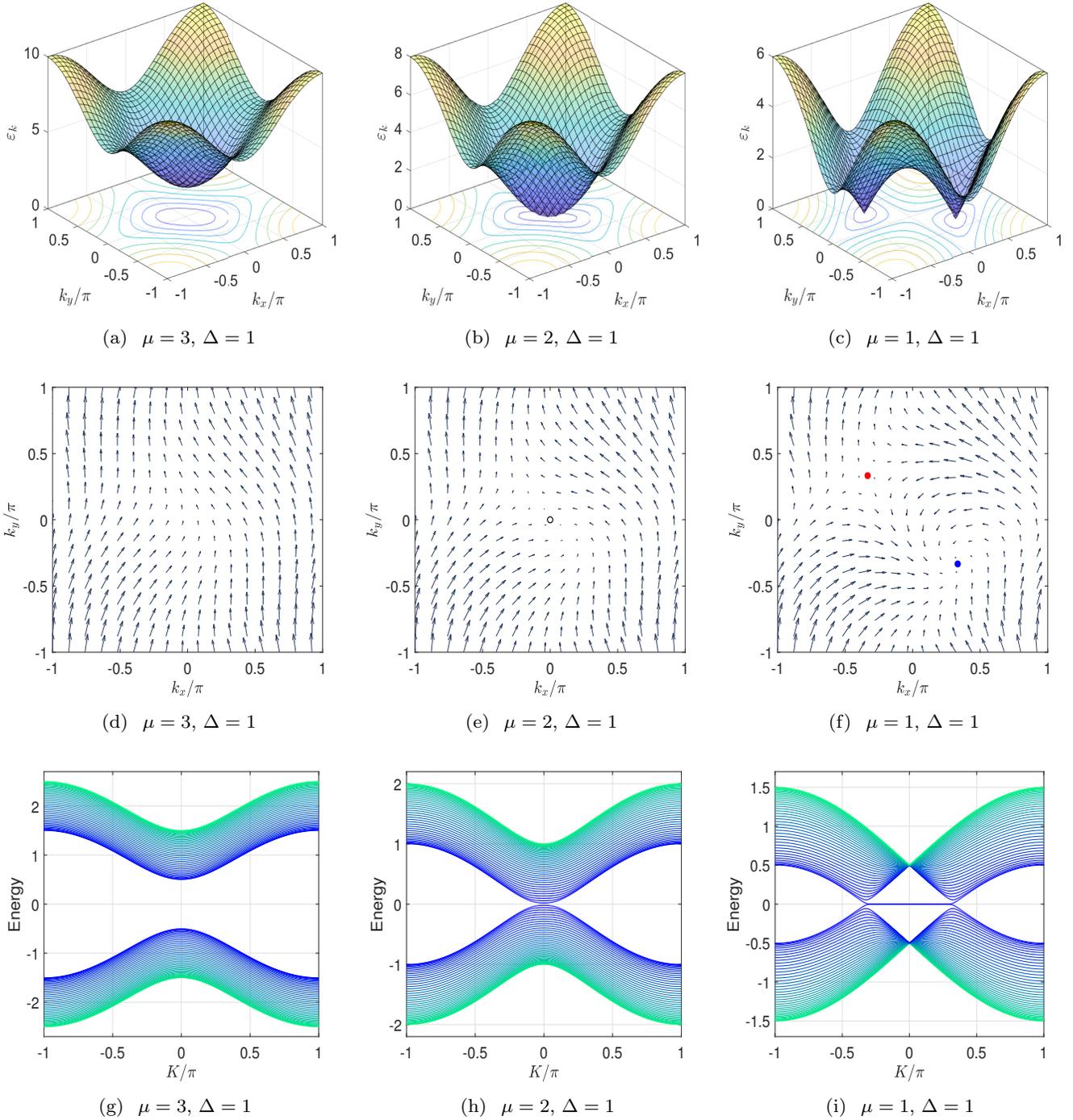}
\caption{(Color online) (a)-(c) Plots of energy spectra from Eq. (\protect
\ref{spectrum}) at three typical points a, b, and c marked in the
phase diagram in Fig. \protect\ref{fig1}.. There is gapped in (a), a single degeneracy
point with parabolic dispersion in (b), and two degeneracy points with linear
dispersion in (c). (d)-(f) Plots of field
defined in Eq. (\protect\ref{field}) in the momentum space for three cases
corresponding to (a)-(c). There are two vortices in (f) with opposite
winding numbers $\pm 1$. As $\protect\mu $\ increases, two vortices close
and merge into a single point in (e). As $\protect\mu $\ increases to 3,
the field become trivial in (d).\ (g)-(i) Plots of
the spectra from Eq. (\protect\ref{spectrum_h}) with $M=40$ for three cases
of (a)-(c). It indicates that the existence of pair of vortices links to a
flat band of Majorana lattice.}
\label{fig2}
\end{figure}

\begin{figure}[h!]
	\centering
	\includegraphics[bb=75 150 530 635, width=1\textwidth, clip]{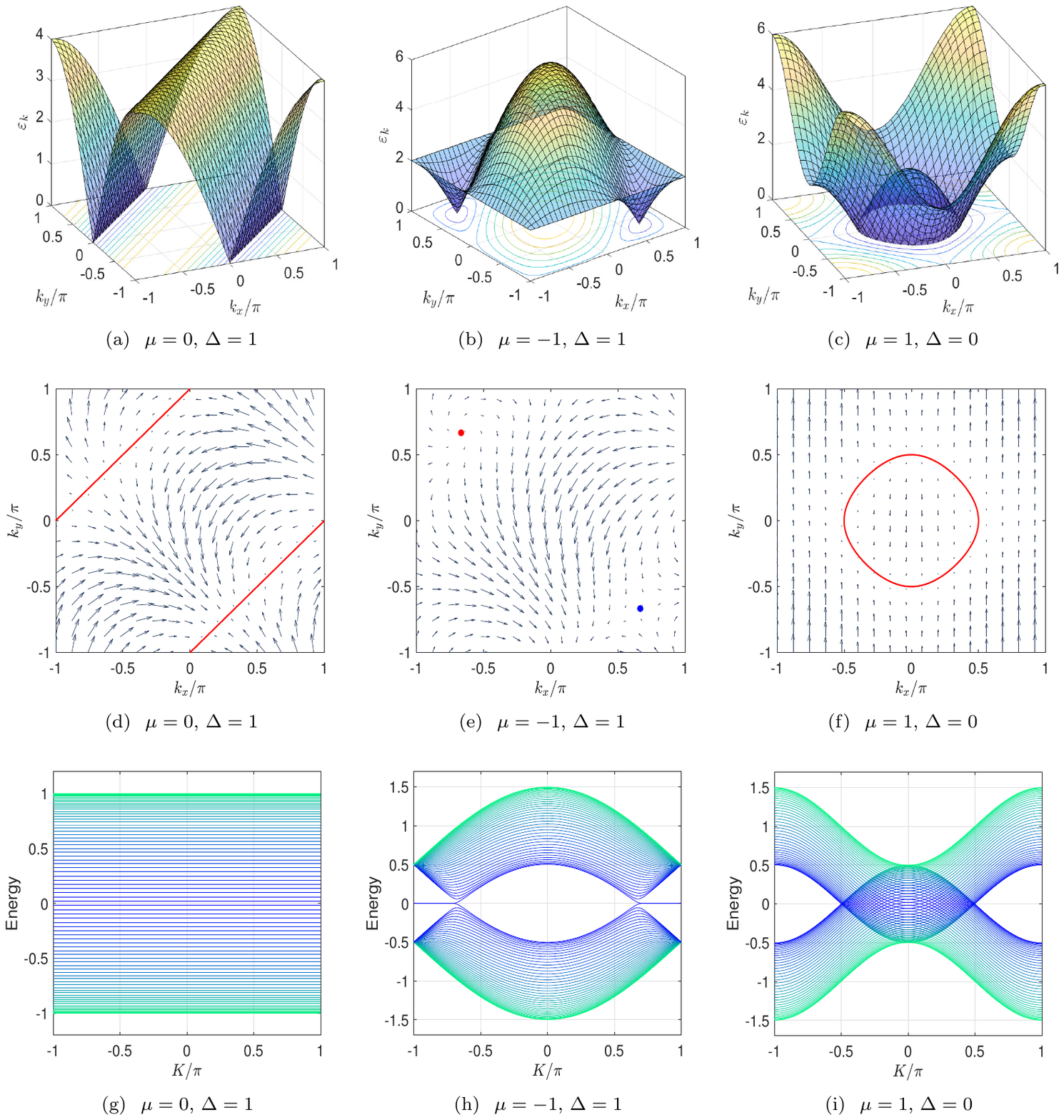}
	\caption{(Color online) (a)-(c) Plots of energy spectra from Eq. (\protect
		\ref{spectrum}) at three typical points d, e, and f marked in the
		phase diagram in Fig. \protect\ref{fig1}.. There is two degeneracy lines in (a), two degeneracy points with linear dispersion in (b), and a degeneracy loop in (c). (d)-(f) Plots of field
		defined in Eq. (\protect\ref{field}) in the momentum space for three cases
		corresponding to (a)-(c). There are two vortices in (e) with opposite
		winding numbers $\pm 1$. (g)-(i) Plots of
		the spectra from Eq. (\protect\ref{spectrum_h}) with $M=40$ for three cases
		of (a)-(c). It indicates that the existence of pair of vortices links to a
		flat band of Majorana lattice.}
	\label{fig3}
\end{figure}

Taking the Fourier transformation 
\begin{equation}
c_{\mathbf{r}}=\frac{1}{N}\sum_{\mathbf{k}}c_{\mathbf{k}}e^{i\mathbf{k}\cdot 
\mathbf{r}},
\end{equation}%
the Hamiltonian with periodic boundary conditions on both directions can be
rewritten as%
\begin{equation}
H=\sum_{\mathbf{k}}\left( 
\begin{array}{cc}
c_{\mathbf{k}}^{\dag } & c_{-\mathbf{k}}%
\end{array}%
\right) h_{\mathbf{k}}\left( 
\begin{array}{c}
c_{\mathbf{k}} \\ 
c_{-\mathbf{k}}^{\dag }%
\end{array}%
\right) ,
\end{equation}%
where%
\begin{equation}
h_{\mathbf{k}}=\left( 
\begin{array}{cc}
\mu -t\cos k_{x}-t\cos k_{y} & i\Delta (\sin k_{x}+\sin k_{y}) \\ 
-i\Delta (\sin k_{x}+\sin k_{y}) & -\mu +t\cos k_{x}+t\cos k_{y}%
\end{array}%
\right) ,
\end{equation}%
where the summation of $\mathbf{k}=(k_{x},k_{y})$ is $\sum_{\mathbf{k}%
}=\sum_{k_{x}=-\pi }^{\pi }\sum_{k_{y}=-\pi }^{\pi }$. The core matrix can
be expressed as%
\begin{equation}
h_{\mathbf{k}}=\mathbf{B}\left( \mathbf{k}\right) \cdot \mathbf{\sigma },
\end{equation}%
where the components of the auxiliary field $\mathbf{B}\left( \mathbf{k}%
\right) =(B_{x},B_{y},B_{z})$\ are%
\begin{equation}
\left\{ 
\begin{array}{c}
B_{x}=0 \\ 
B_{y}=-\Delta (\sin k_{x}+\sin k_{y}) \\ 
B_{z}=\mu -t\cos k_{x}-t\cos k_{y}%
\end{array}%
\right. .  \label{field}
\end{equation}%
$\mathbf{\sigma }$\ are the Pauli matrices

\begin{equation}
\sigma _{x}=\left( 
\begin{array}{cc}
0 & 1 \\ 
1 & 0%
\end{array}%
\right) ,\sigma _{y}=\left( 
\begin{array}{cc}
0 & -i \\ 
i & 0%
\end{array}%
\right) ,\sigma _{z}=\left( 
\begin{array}{cc}
1 & 0 \\ 
0 & -1%
\end{array}%
\right) .
\end{equation}%
The Bogoliubov spectrum is%
\begin{eqnarray}
\varepsilon _{\mathbf{k}} &=&\left\vert \mathbf{B}\left( \mathbf{k}\right)
\right\vert  \notag \\
&=&2\sqrt{\left[ \mu -t\left( \cos k_{x}+\cos k_{y}\right) \right]
^{2}+\Delta ^{2}\left( \sin k_{x}+\sin k_{y}\right) ^{2}}.
\label{spectrum}
\end{eqnarray}%
We are interested in the gapless state arising from the band touching point
of the spectrum. The band degenerate point $\mathbf{k}_{0}=(k_{0x},k_{0y})$
is determined by%
\begin{equation}
\left\{ 
\begin{array}{l}
-\Delta \left( \sin k_{0x}+\sin k_{0y}\right) =0, \\ 
\mu -t\left( \cos k_{0x}+\cos k_{0y}\right) =0.%
\end{array}%
\right.  \label{para eq}
\end{equation}%
As pointed in Ref. \cite{WP1}, two bands touch at three types of
configurations: single point, double points, and curves in the $k_{x}$-$%
k_{y} $\ plane, determined by the region of parameter $\Delta $-$\mu $ plane
(in units of $t$). We focus on the non-trivial case with nonzero $\Delta $.
Then we have%
\begin{equation}
k_{0x}=-k_{0y}=\pm \arccos (\frac{\mu }{2t})  \label{condition eq}
\end{equation}%
in the region $\left\vert \mu /t\right\vert \leqslant 2$, which indicates
that there are two nodal points for $\mu \neq 0$ and $\left\vert \mu
/t\right\vert \neq 2$. The two points move along the line represented by the
equation $k_{0x}=-k_{0y}$, and merge at $\mathbf{k}_{0}=(\pi ,-\pi )$\ when $%
\mu /t=\pm 2$. In the case of $\mu =0$, the nodal points become two nodal
lines represented by the equations $k_{0y}$ $=$ $\pm \pi +k_{0x}$. The phase
diagram is illustrated in Fig. \ref{fig1}, depending on the values of $\mu $
and $\Delta $ (compared with the hopping strength $t$). We plot the band
structures in Fig. \ref{fig2} and Fig. \ref{fig3} for several typical cases.

In the vicinity of the degeneracy points, we have%
\begin{equation}
\left\{ 
\begin{array}{l}
B_{x}=0 \\ 
B_{y}=-\Delta \cos k_{0x}\left( q_{y}+q_{x}\right) \\ 
B_{z}=t\sin k_{0x}\left( q_{x}-q_{y}\right)%
\end{array}%
\right. ,  \label{B field}
\end{equation}%
where $\mathbf{q=k-k}_{0}$, $\mathbf{k}_{0}=\left( k_{0x},k_{0y}\right) $
and $\left( k_{0x},k_{0y}\right) $ satisfy Eq. (\ref{condition eq}), is the
momentum in another frame. Around these degeneracy points, the Hamiltonian $%
h_{\mathbf{k}}$ can be linearized as the form%
\begin{equation}
\mathcal{H}(\mathbf{q})=\sum_{i,j}a_{ij}q_{i}\sigma _{j},
\end{equation}%
which is equivalent to the Hamiltonian for two-dimensional massless
relativistic fermions. Here $\left( q_{1},q_{2}\right) =\left(
q_{x},q_{y}\right) $ and $\left( \sigma _{1},\sigma _{2}\right) =\left(
\sigma _{y},\sigma _{z}\right) $. The corresponding chirality for these
particle is defined as 
\begin{equation}
w=\mathrm{sgn}[\det (a_{ij})].
\end{equation}%
Then we have%
\begin{equation}
\det \left\vert 
\begin{array}{cc}
-\Delta \cos k_{0x} & t\sin k_{0x} \\ 
-\Delta \cos k_{0x} & -t\sin k_{0x}%
\end{array}%
\right\vert =t\Delta \sin 2k_{0x},
\end{equation}%
which leads to $w=\pm 1$ for two nodal points. The chiral relativistic
fermions serve as two-dimensional Weyl fermions. Two Weyl nodes located at
two separated degenerate points have opposite chirality. We note that for $%
\Delta \left\vert \mu /t\right\vert (\left\vert \mu /t\right\vert -2)=0$, we
have $w=0$. At this situation, two Weyl nodes merge at $(0,0)$\ and $(\pm
\pi ,\mp \pi )$. The topology of the nodal point becomes trivial, and a
perturbation hence can open up the energy gap. We illustrate the vortex
structure of the degeneracy point in $k_{x}$-$k_{y}$\ plane in Fig. \ref%
{fig2} and Fig. \ref{fig3}. As shown in figures, we find three types of topological
configurations: pair of vortices with opposite chirality, single trivial
vortex (or degeneracy lines), and no vortex, corresponding to topological
gapless, trivial gapless and gapped phases, respectively.

\subsection{Majorana flat band edge modes}

Now we turn to study the feature of gapless phase in the framework of
Majorana representation. The Kitaev model on a honeycomb lattice and chain
provides well-known examples of systems with such a bulk-boundary
correspondence \cite{AK,GB,DHL,KPS,GK,GK2,GK3}. It is well known that a
sufficient long chain has Majorana modes at its two ends \cite{AYK}. A
number of experimental realizations of such models have found evidence for
such Majorana modes \cite{VM,LPR,AD,ADKF,BA}.\textbf{\ }In contrast to
previous studies based on a gapful system with nonzero Chern number, we
focus on the Kitaev model in the topologically trivial phase. This is
motivated by the desire to get a connection between the Majorana edge modes
and topological nature hidden in a topologically trivial superconductor. At
first, we revisit the description\ of the present model on a cylindrical
lattice\ in terms of Majorana fermions.

We introduce Majorana fermion operators%
\begin{equation}
a_{\mathbf{r}}=c_{\mathbf{r}}^{\dagger }+c_{\mathbf{r}},b_{\mathbf{r}%
}=-i\left( c_{\mathbf{r}}^{\dagger }-c_{\mathbf{r}}\right) ,
\end{equation}%
which satisfy the relations%
\begin{eqnarray}
\left\{ a_{\mathbf{r}},a_{\mathbf{r}^{\prime }}\right\} &=&2\delta _{\mathbf{%
r},\mathbf{r}^{\prime }},\left\{ b_{\mathbf{r}},b_{\mathbf{r}^{\prime
}}\right\} =2\delta _{\mathbf{r},\mathbf{r}^{\prime }},  \notag \\
\left\{ a_{\mathbf{r}},b_{\mathbf{r}^{\prime }}\right\} &=&0,a_{\mathbf{r}%
}^{2}=b_{\mathbf{r}}^{2}=1.
\end{eqnarray}%
Then the Majorana representation of the Hamiltonian is%
\begin{eqnarray}
H&=&\frac{1}{4}\sum_{\mathbf{r}}[i(t+\Delta )\sum_{\mathbf{a}}a_{\mathbf{r}%
}b_{\mathbf{r+a}}  \notag \\
&&+i(t-\Delta )\sum_{\mathbf{a}}a_{\mathbf{r+a}}b_{\mathbf{r}}-i2\mu a_{%
\mathbf{r}}b_{\mathbf{r}}+\text{\textrm{h.c.}}].
\end{eqnarray}%
It represents a honeycomb lattice with extra hopping term $a_{\mathbf{r+a}%
}b_{\mathbf{r}}$, which is schematically illustrated in Fig. \ref{fig1}.
Before a general investigation, we consider a simple case to show that a
flat band Majorana modes do exist.\ Taking $t=\Delta =\mu $ the Hamiltonian
reduces to%
\begin{equation}
H_{\text{hc}}=\frac{t}{2}\sum_{\mathbf{r}}(ia_{\mathbf{r}}\sum_{\mathbf{a}%
}b_{\mathbf{r+a}}-ia_{\mathbf{r}}b_{\mathbf{r}}+\text{\textrm{h.c.}}).
\end{equation}%
which corresponds to a honeycomb ribbon with zigzag boundary condition. It
is well-known that there exist a partial flat band edge modes in such a
lattice system\cite{FM,RS,YW,WK}.

In the following, we will show that this feature still remains in a wide
parameter region. Consider the lattice system on a cylindrical geometry by
taking the periodic boundary condition in one direction and open boundary in
another direction. For a $M\times M$\ Kitaev model, the Majorana Hamiltonian
can be explicitly expressed as

\begin{eqnarray}
H_{\text{M}} &=&\frac{i}{4}\sum_{l,j=1}^{M}[(t-\Delta
)(a_{m,n+1}b_{m,n}+a_{m+1,n}b_{m,n})  \notag \\
&&+(t+\Delta )(a_{m,n}b_{m,n+1}+a_{m,n}b_{m+1,n})-2\mu a_{m,n}b_{m,n}-%
\mathrm{h.c.}],
\end{eqnarray}%
by taking $\mathbf{r=}m\mathbf{i+}n\mathbf{j}\rightarrow (m,n)$. The
boundary conditions are $a_{n,1}=a_{n,M+1}$, $b_{n,1}=b_{n,M+1}$, $%
a_{M+1,n}=0$, and $b_{M+1,n}=0$.

Consider the Fourier transformations of Majorana operators%
\begin{equation}
\left\{ 
\begin{array}{c}
\alpha _{m,K}=\frac{1}{\sqrt{N}}\sum_{n=1}^{M}e^{-iKn}a_{m,n} \\ 
\beta _{m,K}=\frac{1}{\sqrt{N}}\sum_{n=1}^{M}e^{-iKn}b_{m,n}%
\end{array}%
\right. ,
\end{equation}%
where the wave vector $K=2\pi l/N$, $l=1,2,...,N$.\ 

The Hamiltonian $H_{\text{M}}$ can be rewritten as 
\begin{eqnarray}
H_{\text{M}} &=&\sum_{K}h_{\text{M}}^{K},  \label{spectrum_h} \\
h_{\text{M}}^{K} &=&\sum_{m=1}^{M-1}(\eta \alpha _{m,K}i\beta _{m,-K}+\delta
_{1}\alpha _{m+1,K}i\beta _{m,-K}  \notag \\
&&+\delta _{2}\alpha _{m,K}i\beta _{m+1,-K})+\eta \alpha _{M,K}i\beta
_{M,-K}+\mathrm{h.c.,}  \label{SSH}
\end{eqnarray}%
where $\eta =[(t-\Delta )e^{iK}+(t+\Delta )e^{-iK}-2\mu ]/4$, $\delta
_{1}=(t-\Delta )/4$ and $\delta _{2}=(t+\Delta )/4$, and $h_{\text{M}}^{K}$
obeys 
\begin{equation}
\lbrack h_{\text{M}}^{K},h_{\text{M}}^{K^{\prime }}]=0,
\end{equation}%
i.e., $H_{\text{M}}$ has been block diagonalized. We would like to point
that operators $\alpha _{m,K}$ and$\ \beta _{m,K}$ are not Majorana fermion
operators except the cases with\textbf{\ }$K=0$\textbf{\ }or\textbf{\ }$\pi $%
. We refer such operators as to auxiliary\textbf{\ }operators.\textbf{\ }We
note that each\textbf{\ }$h_{\text{M}}^{K}$\textbf{\ }represents a modified
SSH chain about auxiliary operators\textbf{\ }$\alpha _{m,K}$\textbf{\ }and$%
\ \beta _{m,K}$\textbf{\ }with\textbf{\ }$\eta $\textbf{, }$\delta _{1}$%
\textbf{, }and\textbf{\ }$\delta _{2}$\textbf{\ }hopping terms. One can
always get a diagonalized $h_{\text{M}}^{K}$\ through the diagonalization of
the matrix of the corresponding single-particle modified SSH chain. For
simplicity, we only consider the case with positive parameters\textbf{\ }$t,$%
\textbf{\ }$\Delta ,$\textbf{\ }and\textbf{\ }$\mu $\textbf{.} In large $M$
limit, there are two zero modes for $h_{\text{M}}^{K}$\ under the condition $%
0<\mu <2$ (in units of $t$). Actually, it can be checked that\textbf{\ }$h_{%
\text{M}}^{K}$\textbf{\ }can contribute a term 
\begin{equation}
0\times (\gamma _{K}^{\dag }\gamma _{K}-\gamma _{K}\gamma _{K}^{\dag }),
\label{zero term}
\end{equation}%
where%
\begin{equation}
\gamma _{K}=A\sum_{j=1}^{M}\left[ \left( p_{+}^{M-j+1}-p_{-}^{M-j+1}\right)
\alpha _{j,K}+i\left( p_{+}^{j}-p_{-}^{j}\right) \beta _{j,K}\right] .
\end{equation}%
Here $A=\left( 2\sum_{j=1}^{M}\left\vert p_{+}^{j}-p_{-}^{j}\right\vert
^{2}\right) ^{-\frac{1}{2}}$ is normalization constant, and 
\begin{equation}
p_{\pm }=\frac{\pm \sqrt{\left\vert \eta \right\vert ^{2}-4\delta _{1}\delta
_{2}}-\left\vert \eta \right\vert }{2\delta _{2}}.
\end{equation}%
The term in Eq. (\ref{zero term}) exists under the convergence condition%
\begin{equation}
\lim_{j\rightarrow \infty }\left( p_{+}^{j}-p_{-}^{j}\right) =0\mathbf{.}
\label{convergence}
\end{equation}%
As expected, we note that operators\textbf{\ }$\gamma _{K}$\textbf{\ }%
satisfy the fermion commutation relations%
\begin{equation}
\left\{ \gamma _{K},\gamma _{K^{\prime }}^{\dag }\right\} =2\delta
_{KK^{\prime }},\left\{ \gamma _{K},\gamma _{K^{\prime }}\right\} =0,
\end{equation}%
representing edge modes. The sufficient condition for Eq. (\ref{convergence}%
) is 
\begin{equation}
\left\vert p_{\pm }\right\vert <1,
\end{equation}%
which leads to 
\begin{equation}
\left\{ 
\begin{array}{c}
\left\vert \eta \right\vert <2\left\vert \delta _{2}\right\vert  \\ 
-\left\vert \delta _{2}\right\vert ^{2}+\left\vert \eta \delta
_{2}\right\vert <\delta _{1}\delta _{2}\leqslant \left\vert \frac{\eta }{2}%
\right\vert ^{2}%
\end{array}%
\right. ,
\end{equation}%
or more explicitly form 
\begin{equation}
\left\{ 
\begin{array}{l}
R-2t\Delta <0 \\ 
\Delta ^{2}>R+2\Delta ^{2}\geqslant 0%
\end{array}%
\right. ,
\end{equation}%
where $R\left( K,\Delta ,\mu ,t\right) =(t^{2}-\Delta ^{2})\cos ^{2}(K)-2\mu
t\cos (K)+\mu ^{2}-t^{2}$. To demonstrate this result, considering a simple
case with $\Delta =\mu =t,$ we find that $R$\ reduces to $-2\Delta ^{2}\cos
(K)$\ and satisfies above equations when take $-\pi /3<K<\pi /3$.
Furthermore the parameters become $\eta =\Delta (e^{-iK}-1)/2$, $\delta
_{1}=0$ and $\delta _{2}=\Delta /2$, and $h_{\text{M}}^{K}$\ corresponds to
a simple SSH chain with $\left\vert \eta \right\vert <\left\vert \delta
_{2}\right\vert $. The edge mode wave functions can be obtained from $p_{+}=0
$\ and $p_{-}=-\sqrt{2\left( 1-\cos K\right) }$. Then the existence of edge
modes is well reasonable. The zero modes in the plot of energy band in Fig. %
\ref{fig2}(i) corresponds this flat band of edge mode.

\subsection{Disorder perturbation}

\begin{figure}[h!]
\centering
\includegraphics[ bb=75 235 530 555, width=1\textwidth, clip]{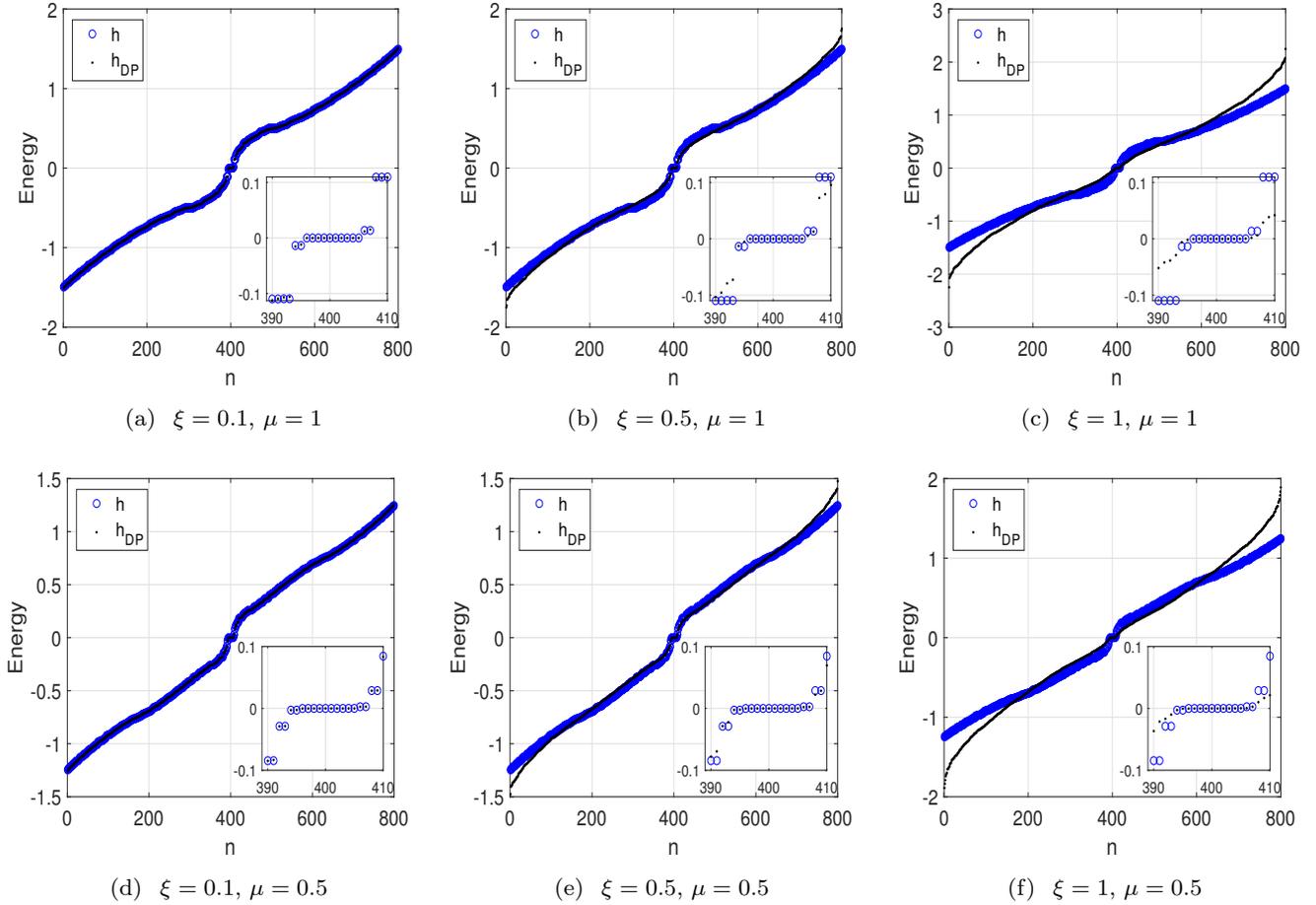}
\caption{(Color online) Plots of eigenvalues of matrices $h$ and $h_{\text{DP%
}}$ with typical parameters and different disorder strength factor $\protect%
\xi $. The dimension of matrix is $800$, corresponding to $M=20$. According
to analytical analysis, there are $10$ and $14$ quasi-zero modes in the
upper and lower panels. Here we take $\Delta=t=1 $. Numerical results show
that as $\protect\xi $\ increases, most of levels of $h_{\text{DP}}$\
deviate from that of $h$, while the zero-mode levels remain unchanged,
indicating the robustness of zero modes against to the disorder.}
\label{fig4}
\end{figure}

One of the most striking features of topologically protected edge states is
the robustness against to certain types of disorder perturbation to the
original Hamiltonian. In this section, we investigate the robustness of the
Majorana edge flat band in the presence of disorder. The disorder we discuss
here arises from the parameters $\left\{ t,\Delta ,\mu \right\} $ in the
Hamiltonian $H_{\text{M}}$. More precisely, one can rewrite the Hamiltonian
in the form 
\begin{equation}
H_{\text{M}}=\psi ^{\dag }h\psi ,
\end{equation}%
where $h$ represents a $2M^{2}\times 2M^{2}$ matrix in the basis%
\begin{eqnarray}
\psi &=&(a_{1,1},ib_{1,1},...,a_{1,M},ib_{1,M},  \notag \\
&&a_{2,1},ib_{2,1},...,a_{2,M},ib_{2,M},  \notag \\
&&...,a_{i,j},ib_{i,j},...,  \notag \\
&&a_{M,1},ib_{M,1},...,a_{M,M},ib_{M,M})^{T}.
\end{eqnarray}%
We introduce the disorder perturbation to $H_{\text{M}}$ by preserving the
time reversal symmetry, i.e., keeping the reality of the parameters $\left\{
t,\Delta ,\mu \right\} $. We take the randomized matrix elements in $h$ by
the replacement%
\begin{equation}
\left\{ 
\begin{array}{c}
t\rightarrow t_{l.j}=r_{l,j}^{a}t \\ 
\Delta \rightarrow \Delta _{l.j}=r_{l,j}^{b}\Delta \\ 
\mu \rightarrow \mu _{l.j}=r_{l,j}^{c}\mu%
\end{array}%
\right. ,
\end{equation}%
to get the disorder matrix $h_{\text{DP}}$. Here $r^{a,b,c}$ are three $%
M\times M$ matrices which consist of random numbers in the interval of $%
\left( 1-\xi ,1+\xi \right) $, influencing each matrix elements. Real factor 
$\xi $\ plays the role of the disorder strength.

We investigate the influence of nonzero $\xi $\ by comparing two sets of
eigenvalues obtained by numerical diagonalization of finite-dimensional
matrices $h$ and $h_{\text{DP}}$, respectively. The plots in Fig. \ref{fig4}.
indicate that the zero modes remain unchanged in the presence of random
perturbations with not too large $\xi $. The numerical result support our
conclusion that the topological gapless states correspond to the presence of
topologically protected edge modes.\newline

\section{Discussion}

According to the bulk-edge correspondence, it seems that the existence of
edge states requires a gapped topological phase. This may not include the
case with a single band which contains topological gapless states. The
topological character of a gapless state does not require the existence of
the gap. This arises the question: What is the essential reason\ for the
edge state, energy gap or topology of the energy band? Obviously, energy gap
is not since many gapped systems do not support the edge states. Then it is
possible that a special single band system supports the edge states. In the
case of the lack of an exact proof, concrete example is desirable. As such
an example we have considered a Kitaev model on a square lattice, which
describes topologically trivial superconductor when gap opens, while
supports topological gapless phase when gap closes. The degeneracy points
are characterized by two vortices, or Weyl nodal points in momentum space,
with opposite winding numbers, which are not removable unless meet together.
We demonstrated that a topologically trivial superconductor emerges as a
topological gapless state, which support Majorana flat band edge modes. The
new quantum state is characterized by two linear band-degeneracy points with
opposite topological invariant. In sharp contrast to the conventional
topological superconductor, such a system has single band, thus has zero
Chern number. We prove that the appearance of this topological feature
attributes to the corresponding Majorana lattice structure, which is a
modified honeycomb lattice. The topological feature of an edge state is the
robustness against disorder. The numerical results indicate that such a
criteria is met for this concrete example. We also note that the topological
gapless state and the edge state have the same energy level, which is also
an open question in the future.

\section*{Acknowledgements}

This work was supported by National Natural Science Foundation of China
(under Grant No. 11874225).

\section*{Author contributions statement}

Z.S. conceived the idea and carried out the study. K.L.Z, P.W. and Z.S.
discussed the results. Z.S and K.L.Z wrote the manuscript with inputs from
all the other authors.

\section*{Additional information}

\textbf{Competing interests:} The authors declare no competing interests.

\end{document}